%% file: main.tex
\documentclass[sigconf]{acmart}

\usepackage{todonotes}
\usepackage{acronym}
\usepackage{csquotes}
\usepackage{csvsimple-l3}

\AtBeginDocument{%
  \providecommand\BibTeX{{%
    \normalfont B\kern-0.5em{\scshape i\kern-0.25em b}\kern-0.8em\TeX}}}

\copyrightyear{2022}
\acmYear{2022}
\setcopyright{acmcopyright}\acmConference[WebSci '22]{14th ACM Web Science Conference 2022}{June 26--29, 2022}{Barcelona, Spain}
\acmBooktitle{14th ACM Web Science Conference 2022 (WebSci '22), June 26--29, 2022, Barcelona, Spain}
\acmPrice{15.00}
\acmDOI{10.1145/3501247.3531567}
\acmISBN{978-1-4503-9191-7/22/06}

\acmConference[WebSci22]{14th ACM Web Science Conference 2022}{June 26--29,
  2022}{Barcelona, Spain}
\acmPrice{15.00}
\acmISBN{978-1-4503-9191-7/22/06}

\setcopyright{none}
\renewcommand\footnotetextcopyrightpermission[1]{}
\setcopyright{none}
\makeatletter
\renewcommand\@formatdoi[1]{\ignorespaces}
\makeatother

\begin{document}

\input{acronyms.tex}

\title[Audition Search Query Bias]{Auditing Search Query Suggestion Bias Through Recursive Algorithm Interrogation}


\author{Fabian Haak}
\email{fabian.haak@th-koeln.de}
\orcid{0000-0002-3392-7860}
\affiliation{%
  \institution{TH Köln -- University of Applied Sciences}
  \city{Cologne}
  \country{Germany}
}

\author{Philipp Schaer}
\email{philipp.schaer@th-koeln.de}
\orcid{0000-0002-8817-4632}
\affiliation{%
  \institution{TH Köln -- University of Applied Sciences}
  \city{Cologne}
  \country{Germany}
}


\renewcommand{\shortauthors}{Haak and Schaer}

\begin{abstract}
Despite their important role in online information search, search query suggestions have not been researched as much as most other aspects of search engines. Although reasons for this are multi-faceted, the sparseness of context and the limited data basis of up to ten suggestions per search query pose the most significant problem in identifying bias in search query suggestions. 
The most proven method to reduce sparseness and improve the validity of bias identification of search query suggestions so far is to consider suggestions from subsequent searches over time for the same query. This work presents a new, alternative approach to search query bias identification that includes less high-level suggestions to deepen the data basis of bias analyses. We employ recursive algorithm interrogation techniques and create suggestion trees that enable access to more subliminal search query suggestions. Based on these suggestions, we investigate topical group bias in person-related searches in the political domain.

\end{abstract}




\maketitle

\section{Introduction}
Search engines are a popular and trusted means of finding information on political topics, not only online information sources considered~\cite{ray_2020, edelman_2017, edelman_2022}. Users communicate their information needs via queries~\cite{belkin_anomalous_1980}, while search query suggestions presented by the search engines to these queries often lead the searches' direction. Although different definitions and types of query suggestions exist, such as query expansions, auto completions, query predictions, or query refinements, we use the term \textquote{query suggestion} to describe suggested search queries that are based on the current input into a search mask of a search engine.
(Search) query suggestions influence the way users search and generate different search engine results, which are known to impact opinion formation and thus public opinions overall \cite{epstein_search_2015, zweig_watching_2017}. 
In an ideal, neutral setting, the influence of query suggestion in the information search process poses no danger. However, especially in opinionated domains such as politics, query suggestions can be opinionated, unrepresentative, or biased towards certain stances, views, or topics. 

Research on the topic of biased query suggestions aims at finding ways to identify biased query suggestions despite the unfavorable conditions posed by the sparse context, non-phrasality, and domain-/topic dependency of query suggestions. 
Identifying suggestions that are problematic due to toxicity, vulgarity, or other forms of negative attribution are caught by the search engines or identified by users most of the time~\cite{google_how_2019}. More problematic are types of bias that arise from biased distributions of topics or other subliminal shifted information exposures. Those cannot be identified from a single list of query suggestions but require larger data sets, which are very rare.

Another problem in analyzing bias in query suggestions arises from how information search usually consists of iterative rephrasing and multiple searches until either the results represent the expected outcome or the information need is constituted and the users are satisfied~\cite{cai_survey_2016}. Since searches are not carried out over a long period, it is not representative to analyze a collection of suggestions that never appeared together at any point in time. Furthermore, query suggestions are dynamically produced based on the user's input. 

This work aims to employ the technique of \ac{RAI}~\cite{robertson_auditing_2019}, previously used to compare the suggestions produced by Google and Bing. In the approach presented in this paper, so-called suggestion trees are used in conjunction with systematic query alteration to retrieve an as-complete-as-possible set of query suggestions. 

Using this methodology, this work attempts to uncover topical group bias in query suggestions for person-related searches with names of politicians as root query inputs. We aim to answer two questions: 
\begin{enumerate}
    \item Are query suggestions for person-related searches in the domain of politics topically biased towards groups?
    \item Does a depth-first suggestion network analysis reveal more or different bias when compared to analyses on a longitudinal \textquote{top-level-suggestion analysis} level~\cite{bonart_investigation_2019} or similar perception-aware approaches~\cite{haak_perception-aware_2021}?
\end{enumerate}

\section{Related Work}
Search engines have a significant impact on shaping political opinion: Epstein and Robertson~\cite{epstein_search_2015} coined the term \ac{SEME} and estimated that the influence of search query suggestions on what users search for could change the outcome of national elections. Although the strength of the \ac{SEME} was questioned~\cite{zweig_watching_2017}, it demonstrated the urgency and seriousness of bias in search engines. Biased search results and thus search suggestions can be problematic, as search engines are attributed with a high level of trustfulness by users~\cite{edelman_2022} that might be exploited. Furthermore, it was shown that query suggestions can be manipulated, and it cannot be told from the suggestions themselves if they are~\cite{wang_2018}. This poses another dangerous aspect of the influence of query suggestions.

Bias can be introduced by the data and algorithms the query suggestions are based on~\cite{l_introna_defining_2000}, or by cognitive user-sided effects~\cite{azzopardi_cognitive_2021}. Search engines display the information inherent in the data; therefore, bias is as pervasive in search results as it is in people's minds and in natural language texts~\cite{bolukbasi_man_2016, noble_algorithms_2018, devlin_bert_2019}. Even Google acknowledges this, stating that they are aware of the partially dis-informative and biased content presented by its search engine~\cite{google_how_2019}. Since query suggestions are based on previous searches, both by user input and interactions with query suggestions, they are highly susceptible for \textquote{destructive feedback loop[s]}~\cite{hiemstra_reducing_2020}, as it was the case for Bettina Wulff, wife of former German President Christian Wulff. In 2012 Wulff undertook legal actions against Google, as searches for her name resulted in direct query suggestions related to \textquote{escort} or \textquote{prostitution}. Due to increased media attention, these searches and their defaming suggestions were searched for and clicked on even more frequently, a classic Matthew effect. Google and other search engine providers have stated that they do filter suggestions to avoid negative effects; still, this only applies to single harmful suggestions
~\cite{yehoshua_google_2016}. 

Algorithm audits are a way to describe and investigate an algorithm's behavior and outcome~\cite{sandvig_auditing_2014}. Auditing approaches have been used to investigate search engines on a larger scale in regards to political bias in terms of partisanship~\cite{hu_auditing_2019}, or personalization effects in political search contexts~\cite{robertson_auditing_2018, le_measuring_2019}.

Biased search results have been discussed and researched. Kulshrestha et al. studied and compared biases in Google search results and Twitter search~\cite{kulshrestha_search_2019}. Their bias detection method relies on peripheral information such as the author, publishing and dissemination platforms, and information derived from these sources. Although numerous systems were proposed to establish an algorithmic fair ranking of search engine results, including an ongoing TREC track~\cite{trec-fair-ranking-2021}, the search suggestion component or autocomplete function of modern web search engines is underrepresented in current research on fairness and bias. 

So far, very few attempts toward detecting bias in search query suggestions have been made. First attempts to uncover bias focused on person-related searches, with a particular focus on politicians~\cite{bonart2018intertemporal}.  
 Later Mertens~et~al.~\cite{mertens_as_2019} and Bonart~et~al.~\cite{bonart_investigation_2019} found a significant bias with regard to gender aspects in searches for politicians. While the first study used a dataset on Twitter searches, the latter analyzed topical bias in search query suggestions to politicians and the development of suggestions over time. They created a dataset by collecting suggestions by longitudinal means, retrieving query suggestions regularly over a span of time, and analyzing the entirety of suggestions for a query~\cite{bonart_investigation_2019}. 
 However, by only analyzing a maximum of ten unique single term suggestions per search, unaware of position and frequency, they do not integrate many of the properties of query suggestions into their research. Other approaches have thus tried to incorporate these properties and how query suggestions are perceived and interacted with. These approaches attempt to detect perceivable bias and introduce new, list-order-aware metrics and frequency counts over time~\cite{haak_perception-aware_2021}. The bias detection pipeline used in these studies was expanded to incorporate the ranking of search suggestions in the autocomplete list with a model of perception-awareness~\cite{haak_perception-aware_2021}, which translates findings on how users perceive and interact with search query suggestions~\cite{hofmann_eye-tracking_2014} to methodology and metrics in bias detection. Using the same dataset, they introduced new metrics that take account of frequency and list position of suggestions, optimizing the validity of the analysis results for inferences for factually perceivable bias. 
Although these studies have much in common with the approach presented in this paper in terms of domain and bias detection approach, they differ significantly regarding the employed dataset. While both previous studies have collected suggestions as longitudinal data over a long period, we analyze cross-sectional suggestion data. 

Other properties of search engine bias were suggested, like dynamic suggestions during search query input, query reformulation strategies, and cognitive effects during the information search process~\cite{azzopardi_cognitive_2021} but only a few of them were systematically examined.

\section{Auditing Search Query Bias}

The methodology presented in this work aims to analyze topical group bias in search query suggestions by extending the data on which the analysis is carried out. Previous approaches to bias identification in search query suggestions mainly suffer from two problems:
\begin{enumerate}
    \item Publicly available data sets are rare and creating search query suggestion datasets by utilizing previously published approaches is time-consuming.
    \item Performing bias analyses on base-level suggestions (suggestions to a root query) alone disregards many suggestions, that a user would be exposed to during a search process while reformulating and iteratively searching for different search queries.
\end{enumerate}
This work aims to overcome these problems. This is achieved by combining the \ac{RAI} approach~\cite{robertson_auditing_2019} and a regression-based topical search query suggestion pipeline similar to~\cite{bonart_investigation_2019, haak_perception-aware_2021} in a bias detection framework.

\begin{figure}[t]
  
  \centering
  \includegraphics[width=\linewidth]{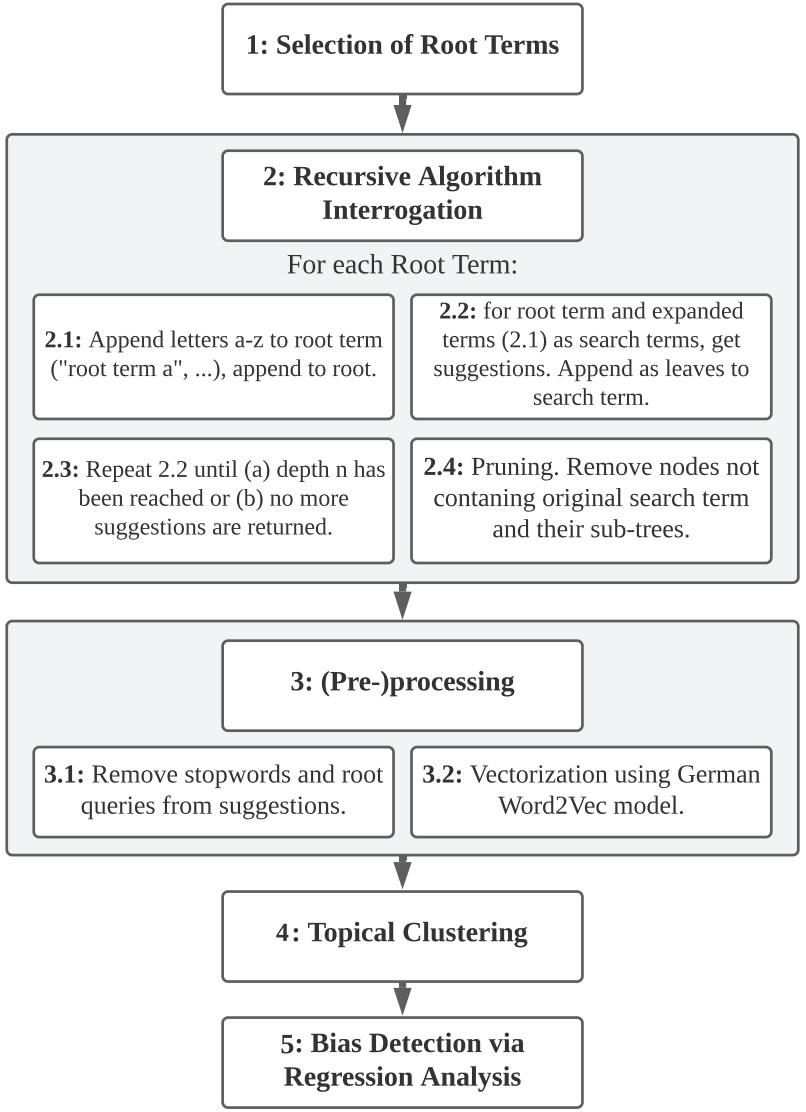}
    
  \caption{Schematic view of the methodological approach for RAI-based bias detection using suggestion trees introduced by~\cite{robertson_auditing_2019}.}
  \Description{Schematic view of the methodological approach for RAI-based bias detection using suggestion trees.}
\label{fig:flow}
\end{figure}

The components and steps of the methodological approach are visualized in \autoref{fig:flow}. First, root terms are selected. Next \ac{RAI} is used to create a data set consisting of search query suggestions for the selection of root queries. The resulting data is preprocessed, the processed suggestions are vectorized, and clustered topically. Finally, group bias is analyzed by performing regression analyses.

\begin{figure*}[t]

  \centering
  \includegraphics[width=\linewidth]{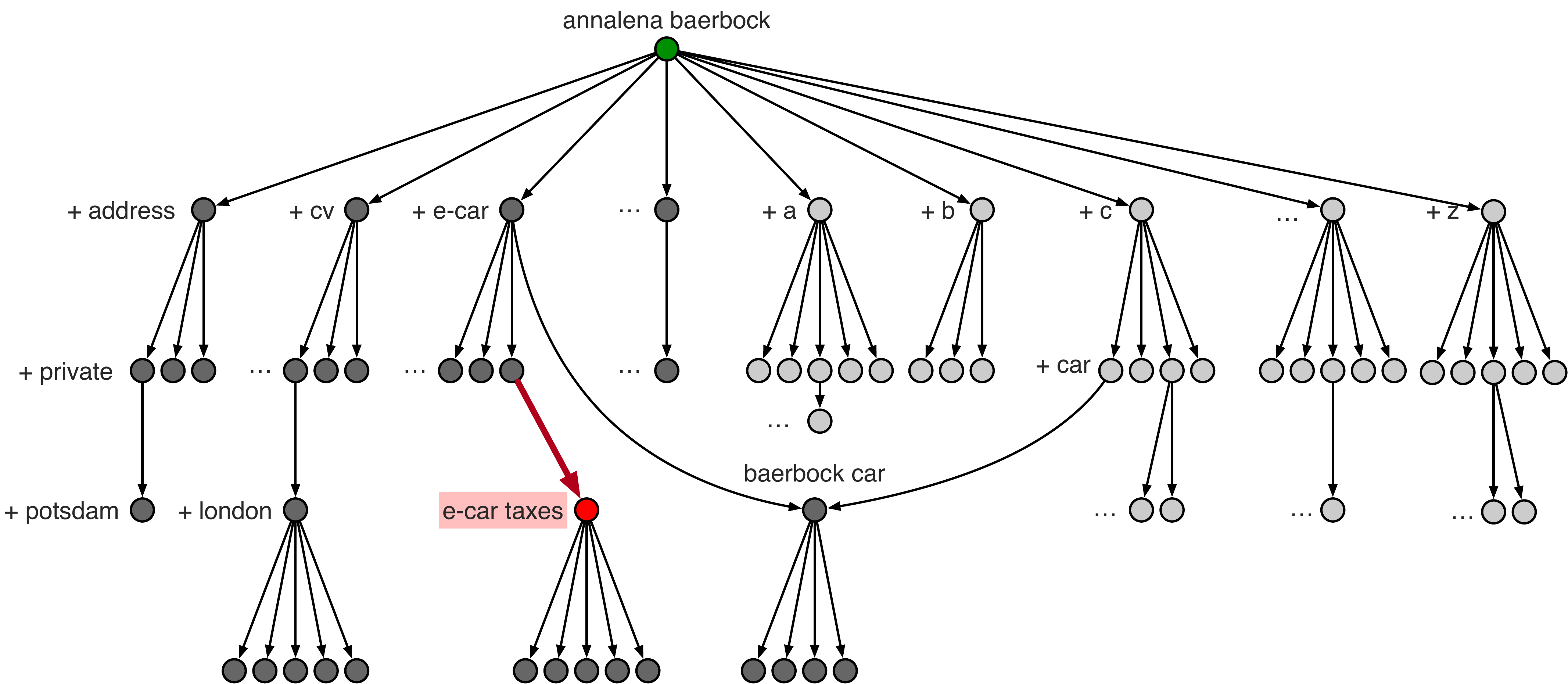}

  \caption{Example for a suggestion tree based on the root term \textquote{annalena baerbock} (the German Federal Foreign Minister, former co-leader of the Green Party), marked in green. Suggestion terms added to a former query string are marked with~+; otherwise, the substituted query string is entirely written out. In order to preserve the clarity of the figure, some labels have been omitted. The dark grey nodes show suggested terms based on the former query. Light grey nodes are suggestions based on a first letter suggestion or their subsequent suggestions. Nodes with suggestions unrelated to the initial root query are marked in red. These nodes are pruned, including their following sub-trees.}
\label{fig:sug_tree}
\end{figure*}

\subsection{Query Suggestion Extraction with Recursive Algorithm Interrogation} \label{chap:rai}
The following section gives an overview of the method of using \ac{RAI} for creating suggestion trees. 

First, a list of root queries is assembled.
For each root query, obtaining a suggestion tree with the term as a root starts with retrieving the query suggestions for a root query (usually up to ten) and adding them to leaves with the original term as their parent. Then, for each of these suggestions as new input queries, suggestions are crawled and added to the tree as child nodes of the previous suggestion. This is repeated recursively until a tree with depth n is created or if there are no more suggestions for any of the terms at the current level. This produces breadth-first-networks of suggestions, compiled in tree structures (suggestion trees). 
In addition to the \ac{RAI} method originally presented, for our approach, each of the search terms is expanded by appending each letter a to z to the end of the query. This procedure widens the first layer of the search tree by starting with up to 270  suggestions for the first layer alone (26*10 plus 10 for the root query), compared to at max ten suggestions produced for the root query alone. Our goal is to find suggestions that a user could be exposed to when entering a query.
The large amount of suggestions illustrates a significant problem with previous approaches to detecting bias in search query suggestions. Analyzing a maximum of up to ten suggestions for each search term disregards the iterative nature of online information search and suggestions during user input. 
The next step consists of pruning the suggestion tree. Leaves that are not semantically connected to the root query are removed along with their complete sub-trees. For example, the input query \textquote{volker wissing}, the current federal minister for digital and transport, one suggestion is \textquote{wissing}, which is not only the name of a politician but also the name of a city. Suggestions for \textquote{wissing} as a new input query do not relate to the person Volker Wissing, but the city. We assume that a user would not interact with this particular suggestion when trying to find information on Volker Wissing. Therefore very likely, neither of those suggestions is relevant. \autoref{fig:sug_tree} shows another example for pruning, highlighted red. Here, \textquote{e-car taxes} is returned as a suggestion, and it along with the sub-tree it spans are removed. Pruning these sub-trees inhibits these suggestions from distorting the bias detection results.

Suggestion trees were initially used to compare the suggestions produced by the search engines Google and Bing and describe how both search engines behave on deeper suggestion levels~\cite{robertson_auditing_2019}. Their findings imply that Google produces an increased amount of and more diverse queries than Bing. While bias analysis was never their goal, \ac{RAI} offers advantages over previous methods for creating query suggestion data sets.
Previous approaches gathered base-level suggestions for query terms repeatedly, often over a long period of time~\cite{bonart_investigation_2019}. This has the advantage that it is possible to investigate how suggestions change over time. However, by omitting all suggestions except for the ten top-level suggestions, most suggestions a user could and would be exposed during information search for any query other than the root term, for example, by interacting with one of the suggestions, are not taken into account. Doing so thus compromises the validity of the results of bias analyses conducted in this way. Furthermore, if the root queries are from a very homogeneous selection and have many prominent common traits, their base-level suggestions tend to be similar to another. For example, when crawling suggestions to German politicians, many suggestions are similar because they are very general or relate to the office the person occupies rather than personal characteristics (such as \textquote{age}, \textquote{twitter}, \textquote{husband/spouse}, \textquote{election} or \textquote{income}). 
Another advantage of using \ac{RAI} is that it makes the time-intensive crawling suggestions to many queries to construct a data set virtually obsolete. Overall, including the full spectrum of query suggestions for a given root query by the means of \ac{RAI} and suggestion tree parsing thus is a promising solution to this problem. 
Finally, the approach to data collection is the only decisive difference from our method. Therefore, by comparing the findings of our study with the findings of these studies, we can directly investigate the influence of the data collection on bias findings in query suggestions. Furthermore, since we are trying to reflect average exposure to suggestions during an iterative search session, we can judge whether bias in such an iterative exposure differs from bias in suggestion exposure to suggestions to only the root query.

\subsection{Methodological Approach}
\label{chap:methodoloy}
The structure and sequence of components within the methodological approach are visualized in \autoref{fig:flow}.

\subsubsection{Dataset Creation and Preprocessing}

\paragraph{Selection of Root Terms and Suggestion Crawl} Since bias towards (groups of) persons has been deemed especially problematic, we chose names as root queries. To keep the results comparable to the results presented in the previously mentioned works, we focus on the political domain and use a selection of names of German politicians as root queries. Names of the federal ministers of Germany and their predecessors, the current prime ministers of the federal states and leaders of the seven German parties with the most votes in the 2021 federal election as well as their predecessors have been selected for a total of 55 root terms. For all of these, demographic information on the meta-attributes gender, party affiliation, political role, and year of birth have been retrieved from the federal returning officer website for the German federal elections~\cite{bwl}. Although the number of root terms is comparably low, the distribution of the meta attributes reflects that of high-level German politicians in general. While about 69 percent of the members in the German federal parliament are male \cite{bwl}, 35 of the 55 or about 64 percent of the politicians that serve as our root queries are male. Similarly, the distribution of party affiliation is very similar to that of the federal parliament.
Then, 8-depth suggestion trees are created using the \ac{RAI} procedure as described in \autoref{chap:rai}. The method had to be adjusted to enable the collection of query suggestions in German and support umlauts. Suggestions were crawled from Google over the span of two days in February 2022. We choose to focus our research on Google to ensure our findings are comparable to previous studies that also used Google suggestions in their data sets. Crawling resulted in a total of 21,407 suggestions in total.   

\begin{table}[t]
    \centering
    \begin{tabular}{c|c}
    preprocessing step & unique suggestions\\\hline
      RAI crawl   & 8,899 \\\hline
      Pruning   & 8,689 \\\hline
      removing ambiguous & 8,079\\
      root term & \\\hline
      removing suggestions & 7,919 \\
      without vector representation &
    \end{tabular}
    \caption{The number of unique suggestions in the data set at steps of the preprocessing progress.}
    \label{tab:corpus_size}
\end{table}

\begin{table*}[t]
\caption{Results of regression analysis of suggestion data from Google. Only meta-attributes with significant differences were included and significant effects ($p \leq \alpha = 0.05$) have been highlighted. }
\begin{tabular}{l|ll|ll|ll|ll}
                & \multicolumn{2}{l|}{Cluster 1:}             & \multicolumn{2}{l|}{Cluster 2:}  & \multicolumn{2}{l|}{Cluster 3:}  & \multicolumn{2}{l}{Number of}   \\
                & \multicolumn{2}{l|}{Political} & \multicolumn{2}{l|}{Locations}   & \multicolumn{2}{l|}{Personal}     & \multicolumn{2}{l}{Suggestions} \\ \hline
                & R2                    & p                   & R2              & p              & R2              & p              & R2              & p             \\
Gender: female  & \textbf{0.074}        & \textbf{0.045}      & \textbf{-0.096} & \textbf{0.022} & 0.003           & 0.702          & -0.026          & 0.235         \\
Minister        & \textbf{0.106}        & \textbf{0.015}      & 0.001           & 0.8            & \textbf{-0.092} & \textbf{0.025} & 0.01            & 0.473         \\
Former Minister & -0.043                & 0.129               & -0.027          & 0.234          & \textbf{0.103}  & \textbf{0.017} & 0.04            & 0.141         \\
CSU             & \textbf{-0.071}       & \textbf{0.049}      & 0.003           & 0.705          & 0.031           & 0.201          & 0.056           & 0.082        
\end{tabular}
\label{tab:results_regression}
\end{table*}

\paragraph{Preprocessing: Cleaning data} Next, preprocessing removes unnecessary and problematic data from the data set. First, the root terms (names of politicians) are removed from all queries except for the root query. This unifies the query suggestions and enables the comparison of the trees with each other. Parts of queries, that are reformulated on subsequent levels of suggestions (e.g. \textquote{address} and \textquote{private} in the sub-tree \textquote{(annalena baerbock) address} $\xrightarrow{}$ \textquote{(annalena baerbock) address private} $\xrightarrow{}$ \textquote{(annalena baerbock) address private potsdam}, compare \autoref{fig:sug_tree}) are not removed from the query, because we assume, that a deeper sub-tree reflects users' interest in this topic and the reformulated words are essential to meaning of the query. 
After removing the root queries, 8,899 unique suggestions remain, with on average 307 unique suggestions per root term, ranging from a minimum of 48 (\textquote{andreas bovenschulte}) to a maximum of 800 (\textquote{olaf scholz}). \autoref{tab:corpus_size} shows the number of suggestions in the data set after each preprocessing step. Next, pruning is performed as described in \autoref{chap:rai}. It removes all sub-trees whose root does not contain the original root term. To account for misspelling and umlaut variations, we manually created a list of variations for each root term that do not get pruned. Combined, pruning of all suggestion trees removes 210 suggestions from the Google data set. 

Aside from root terms, stopwords were removed, and the resulting suggestions without stopwords were saved separately. Reviewing the produced suggestion trees revealed a problem with one of the root terms. \textquote{Gerd Müller} is a former federal minister, but there is a better-known soccer player with the same name. The data indicated that most suggestions related to the athlete instead of the politician, and due to the sparseness and topical ambiguity of most suggestions, we decided to drop the tree and all connected suggestions. This reduced the number of suggestions in the data set to 8,079 (compare \autoref{tab:corpus_size}).

\paragraph{Preprocessing: Vectorization}\label{chap:vec} For semantic representations of the suggestions, vectorization of these suggestions is performed using a large Word2Vec model trained on the CoNLL 2017 corpus~\cite{CoNLL_2017_corpus} and provided by~\cite{fares_word_2017}. A range of different vector models with varying levels of word coverage, complexity, and data basis were compared along with clustering approaches. The resulting clusters were then judged manually for their inner topic cohesion. Based on these judgments, the Word2Vec has proven to produce the most qualitative clusters: It was easy to identify a descriptive quality that spans each cluster as the produced clusters are topically disjunct. 
For suggestions with more than one term, the mean of the vectors of the terms was used as a sentence representation. Although this approach can lead to sub-optimal results for query suggestions that are not sentences but rather collections of keywords, it performed well.
We decided to omit suggestions with tokens for which the vector model does not contain representations. However, this was only the case for a small percentage of suggestions, most of which consist of terms around the coronavirus, for example, \textquote{coronaidiot} or \textquote{corona-test} or misspelled terms. 
After this last preprocessing step, a total of 7,919 unique suggestions from all trees were aggregated into the data set used in the bias analysis.

\begin{table}[t]

    \centering
    \caption{Examples for terms of the k-means clusters, translated to English.}
    \begin{tabular}{c|c|c}
      Cluster 1:  & Cluster 2: & Cluster 3: \\ 
      Politics & Locations & Personal \\ \hline
      ministry of labor  & Berlin & hair dresser\\
      federal minister& Dortmund address & divorce wife\\
      nato russia & Aachen university & tv appearance\\
      \dots & \dots & \dots\\
    \end{tabular}

    \label{tab:clusters}
\end{table}

\subsection{Semantic Clustering}
Since the goal is to compare topics within the query suggestions for each politician to identify biased topical associations of groups of meta-attributes (gender, political party, age), we need to assign the query suggestions to topical clusters. This is achieved by performing a k-means clustering. The evaluation of vectorization- and clustering approaches presented in \autoref{chap:vec} showed that k-means produced better clusters than LDA, DBSCAN, or simple similarity measure-based clustering approaches. The ideal number of clusters was determined to be three by using the sum of squared distances within each cluster for one to ten clusters. An average silhouette analysis supports this and shows the highest silhouette score for three clusters~\cite{ROUSSEEUW198753}. Manually comparing the content of the clusters for three, four, and five clusters also shows, that for the latter two, there are one or two pairs of clusters, that essentially contain the same type of suggestions, further indicating that performing k-means clustering approach on the given set of query suggestions finds three main categories in the data. Furthermore, having three clusters facilitates comparing our results to previously mentioned studies on bias in search query suggestions to person-related search in the political domain, which used very similar clusters \cite{bonart_investigation_2019, haak_perception-aware_2021}. 

\paragraph{Resulting Clusters} The resulting clusters were given descriptive names by studying the terms that have been assigned to each of them. \autoref{tab:clusters} shows some example terms for the clusters. The first cluster can be labeled as political terms. They range from directly containing political terms (\textquote{supreme court debate}, \textquote{minister email}, \textquote{election 2022}, \textquote{government spokesperson}) to suggestions with economical or professional context (\textquote{automotive industry}, \textquote{policemen}, \textquote{attorney}). Suggestions that have been assigned the second cluster have in common that they contain names of locations like cities or countries (\textquote{prof. leipzig anatomy}, \textquote{architect bochum}, \textquote{aachen university}). Despite this shared trait, suggestions of cluster 2 have nothing else in common and could also fit one of the other two clusters (\textquote{lennestadt passed away}, \textquote{election berlin}). The third cluster consists of terms that can be described as personal. Many suggestions contain family- or relationship related terms such as \textquote{wife} or \textquote{children}, others relate to appearance (\textquote{hair}, \textquote{beautiful}, \textquote{has a glass eye}) or are not German (not translated: \textquote{freedom day}). The large variety of topics within the cluster could be problematic for any other domain. Since politicians usually appear very professional and thoughtful (for example, politicians rarely talk about non-politics-related topics publicly), we deem distinguishing between political and non-political as sufficient. Suggestions of the second cluster could often be assigned to either of the other clusters. Due to the many location suggestions and their navigational nature, a separate location cluster improves the validity of comparing the other two clusters by reducing noise.

\subsection{Topical Bias Analysis} 

On a neutral basis, for a leading politician, gender, age, and to a certain degree political affiliation of a politician should not influence what aspects of that politician interests people. Therefore, we would expect no difference in query suggestion between politicians of different groups, and thus any significant differences would indicate bias. Using the topically clustered suggestions, we can investigate whether query suggestions for different groups of politicians with common meta-attribute traits have different topical distributions and are therefore subject to systematic topical group bias. The groups of politicians we focus on are defined by the meta-attributes of gender, age, party affiliation, and political role. We investigate a type of group bias referred to as gender bias by investigating whether there is a difference between suggestions for female and male politicians. Another attribute is the politician's age, based on the year of birth. Furthermore, the party affiliation is factored in the bias analysis, depending on what political party the person is a member of. Their affiliation to one of the German parties SPD, CDU, CSU, FDP, AFD, The Left, or Alliance 90/The Greens is specified in this category. Finally, political role describes due to what group affiliation the person was added to the list of root terms. Categories for this meta-attribute are federal ministers for the 2017 and 2021 legislative period, current prime minister of federated states, and leaders of one of the aforementioned German parties.

The goal is to detect group bias as distinctions in topical exposure between the different groups of meta-attributes. To reveal significant differences, linear regression is performed. Dichotomous dummy variables for the meta-attributes serve as independent variables, while percentages of suggestions of each of the clusters for each politician are used as dependent variables. The model of this regression for topical clusters $c\in1,...,k$  can be expressed as
\begin{equation}
    y_{i,c}=\beta_0+\beta_1x_{i,1}+...+\beta_px_{i,p}+\epsilon_i,
\end{equation}
where $\epsilon_i$ is the independent error term and $i=1,...,N$ are the observation indices. A significant coefficient $\beta_p$ shows a topical bias for this cluster and the attribute $x_{i,p}$. To avoid multicollinearity, one variable per attribute is used as the base category and omitted. 

\paragraph{Findings on Bias in Political Query Suggestions} 
An overview of the significant findings of the regression analysis carried out on the Google query suggestions can be found in \autoref{tab:results_regression}. Despite the unprecedented completeness of query suggestions retrieved by RAI, we did not find a more substantial bias towards the groups of meta-attributes we focused our research on than in previous studies. 

Performing a linear regression analysis with gender as the only explanatory variable shows significant bias. The regression analysis results indicate a gender bias for two of the three clusters. Most notably, they reveal a topical gender bias, showing a significantly different percentage of query suggestions related to professional and political topics depending on the gender of the person whose name was used as a root query. Interestingly, on the given dataset, searches for female politicians return more suggestions that have a political or economic topic than for male politicians ($0.045 = p \leq \alpha = 0.05$). The R-squared-measure indicates that the gender of the politicians can explain 7.4 percent of the difference in percentages of suggestions of the cluster of political suggestions. More political suggestions for names of female politicians is the opposite of the results of base-level-centered bias analyses, which have found significantly fewer political suggestions for women  (compare~\cite{bonart_investigation_2019, haak_perception-aware_2021}). 

Suggestions of the cluster that include names of locations also appear unequally between male and female politicians. Query suggestions for female politicians contain significantly less suggestions that can be attributed to this cluster ($0.022 = p \leq \alpha = 0.05$). The difference between the genders explains 9.6 percent of the variance in the percentages of suggestions of cluster 2. 

Aside from gender, the political role was a biasing factor in the topical distribution of query suggestions. Current ministers of Germany are presented with significantly more political and economical cluster 1 suggestions ($0.015 = p \leq \alpha = 0.05$), and significantly less suggestions that have been assigned to the cluster containing personal query suggestions ($0.025 = p \leq \alpha = 0.05$). The regression results suggest that the group affiliation minister explains 10.6 percent of the variance in the data of cluster 1 and 9.2 percent of the variance in cluster 3.

Age and affiliation to most parties did not reveal to be biased factors in terms of the topical distribution of the three topic clusters investigated in this work. Only politicians of the CSU have been suggested significantly fewer political query suggestions than politicians of other parties ($0.049 = p \leq \alpha = 0.05$). Since previous studies on topical bias in query suggestions employed very similar clusters with comparable cluster descriptions~\cite{bonart_investigation_2019, haak_perception-aware_2021}, we expected to find similarly biased results when comparing the topical distributions in the query suggestions of the same groups of meta-attributes used in these studies. In general, we found similar numbers of cases of bias. The identified gender bias, in stark contrast to the findings of previous studies, we found to be the opposite of what previous studies indicate in terms of bias for politics- and economics-related suggestions. 
The total number of suggestions seems to be unbiased for all groups of meta-attributes.
The overall small coefficients could be a result of the unbiased nature of query suggestion in the political domain; other studies have not found more significant bias.

\section{Discussion}
Despite the low amount of bias discovered in query suggestions for names of German politicians, the approach of employing \ac{RAI} for dataset creation in bias detection was very successful. With little effort, the technique produced large amounts of data that enable researchers to analyze the \textquote{complete picture} of bias users can be exposed to when searching for a person. Especially expanding the root query by appending letters has proven to be an adequate means of efficiently gathering most if not all of the suggestions for a root query.

The results of our study partially contradict previous findings on group bias in search query suggestions for person-related searches in the political domain. Instead of getting more personal suggestions, we identified more political suggestions for female politicians. The only factors of the approach that is significantly different from these previous studies are the inclusion of n-grams (both~\cite{bonart_investigation_2019} and~\cite{haak_perception-aware_2021} did only use single term suggestions for detecting bias) and cross-sectional data instead of longitudinal suggestion data. The consequences of including n-grams can only be assumed. An advantage of using only single term suggestions is that the utilized methods for topical clustering produce much better results. In contrast, not including n-grams ignores the majority of suggestions and compromises the validity of any findings on bias severely. Assuming n-grams are topic-independent, we conclude that topical group bias in this domain varies between base-level suggestions and comparatively rarer suggestions on lower suggestion levels.

Bias analysis of query suggestions poses unique challenges compared to other types of documents. While one of the major challenges lies in the missing context, another important aspect is how query suggestions are perceived.
The introduction of perception-awareness in the form of rank- and frequency-aware factors raised the validity of results of topical bias analysis of search query suggestions~\cite{haak_perception-aware_2021}. Although this work had the similar goal of increasing the validity, we neither did incorporate rank- nor frequency awareness directly. Integrating frequency in the way previous approaches did was no option because the data set was gathered in a single crawl. However, by terms repeating in subsequent suggestions to suggestions (\textquote{karl lauterbach girlfriend} $\longrightarrow$ \textquote{karl lauterbach ex girlfriend}), the frequency of terms is taken into account. Rank-aware measures are an important means of differentiating the perceptibility of suggestions at the different positions in the list of query suggestions. Although we could have adopted rank-aware measures, we decided to use set-based measures. Since we did not have any statistical evidence for weighing the perceptibility of suggestions at depth at or close to the root query compared to suggestions further down the suggestion tree, there was no way to integrate rank weights effectively. 

Since this work did not support previous findings on topical group bias in query suggestions in the political domain, further studies are required to evaluate the presented approach's effectiveness for topical bias detection in search query suggestions. The contradiction in the results of this and previous studies can be attributed to different reasons: Either the current best practice approach of clustering and regression analysis is not suited well enough for the sparse nature of non-phrasal query suggestions, or bias in base-level suggestions significantly differs from the complete set of query suggestions related to the root query term. Since most studies on bias in search query suggestions have been carried out in the political domain, the limited findings could also be attributed to the domain.

\paragraph{Outlook}
Since, despite remaining uncertainties, the approaches presented in this work are very promising, follow-up studies are already being planned. These could leverage the advantages of \ac{RAI} further by employing more techniques to create large, \textquote{perception-aware}\cite{haak_perception-aware_2021} datasets. For example, one possible approach would be to integrate agent-based algorithm audits or user-generated query logs into the method mix.
We see the main weakness of our approach in the approaches to vectorization and topical clustering we used. Although it was beneficial for comparing the results of our study to those of previous studies, using word embeddings trained on a corpus of texts substantially different to search queries is problematic. In the case of our study, the amount of clustered suggestions was not very large, and we were able to ensure, through qualitative analysis, that clustering produced decent results despite sub-optimal word embeddings. Furthermore, using more sophisticated clustering approaches than k-means could reduce noise and improve the overall validity of our results.
The omission of rank-aware factors and measures in our analysis can be attributed to a lack of insight on how to weight suggestions at different depths of the suggestion tree. This could be further investigated with user studies and integrated into the corpus creation or by weighting suggestions at different ranks differently.
Applying the technique to more search engines and other domains could help investigate the approach's suitability for detecting bias in search query suggestions and gaining more insight into the subject. In addition, investigation of biases in search suggestions from different search engines could be a valuable aspect of follow-up studies.

\paragraph{Conclusion}
In this work, we present a new approach to search query bias identification. Our approach combines state-of-the-art techniques for suggestion parsing and proven approaches to topical bias detection in query suggestions. The goal was to enhance the validity of bias analyses for query suggestions by incorporating more complete representations of query suggestions to search terms. 

By methodically crawling base-level suggestions to a root query, expanding the root query, then crawling suggestions recursively for all of these up to a certain depth, we try to create a dataset that includes virtually all suggestions a user could be confronted with while searching for information on the root query. 
As it is best practice for bias identification for query suggestions, we then used clustering and regression analysis to find any biased patterns in the distribution of topics in query suggestions for politicians of certain groups of meta-attributes (e.g., gender). 

The results show some instances of group bias. First, a topical gender bias identified significantly more suggestions that can be described as political for female politicians. Furthermore, the political role of politicians was found to be a biasing meta-attribute. Current ministers of the German government have a significantly higher percentage of political suggestions and fewer personal suggestions. However, searches for former ministers were presented with significantly more personal suggestions. 
These findings differ significantly from other studies on bias in search query suggestions to person-related terms in the political domain. Therefore, we assume that including a complete set of query suggestions instead of just the ten base-level suggestions for each root term causes this difference.


  \bibliographystyle{ACM-Reference-Format}
  \bibliography{references}

\appendix

\begin{table*}[t]
\label{tab:politicians}
    \centering
    \caption{Politicians that serve as root query terms, with their metadata.}

\begin{filecontents*}{final_pol_data.csv}
name,google-suggestions,cluster-0,cluster-2,cluster-1,gender,female,year-of-birth,party,pol-role
alexander gauland,183,0.268,0.678,0.055,male,0,1941,AfD,former party leader
andreas bovenschulte,48,0.188,0.708,0.104,male,0,1965,SPD,minister-president
andreas scheuer,270,0.27,0.637,0.093,male,0,1974,CSU,former minister
angela merkel,602,0.204,0.748,0.048,female,1,1954,CDU,former minister
anja karliczek,140,0.257,0.686,0.057,female,1,1971,CDU,former minister
annalena baerbock,654,0.229,0.734,0.037,female,1,1980,GRÜNE,minister
anne spiegel,229,0.358,0.568,0.074,female,1,1980,GRÜNE,minister
annegret kramp-karrenbauer,250,0.288,0.652,0.06,female,1,1962,CDU,former minister
armin laschet,633,0.3,0.616,0.084,male,0,1961,CDU,former party leader
bettina stark-watzinger,88,0.33,0.614,0.057,female,1,1968,FDP,minister
bodo ramelow,259,0.29,0.602,0.108,male,0,1956,DIE LINKE,minister-president
cem özdemir,320,0.338,0.616,0.047,male,0,1965,GRÜNE,minister
christian lindner,447,0.221,0.685,0.094,male,0,1979,FDP,minister
christine lambrecht,219,0.347,0.612,0.041,female,1,1965,SPD,minister
daniel günther,325,0.243,0.566,0.191,male,0,1973,CDU,minister-president
dietmar woidke,73,0.342,0.616,0.041,male,0,1961,SPD,minister-president
franziska giffey,261,0.295,0.636,0.069,female,1,1978,SPD,former minister
friedrich merz,458,0.365,0.563,0.072,male,0,1955,CDU,party leader
gerd müller,742,0.102,0.756,0.142,male,0,1955,CSU,former minister
heiko maas,439,0.339,0.629,0.032,male,0,1966,SPD,former minister
helge braun,310,0.332,0.603,0.065,male,0,1972,CDU,former minister
hendrik wüst,197,0.33,0.548,0.122,male,0,1975,CDU,minister-president
horst seehofer,268,0.25,0.683,0.067,male,0,1949,CSU,former minister
hubertus heil,286,0.357,0.549,0.094,male,0,1972,SPD,minister
janine wissler,231,0.364,0.528,0.108,female,1,1981,DIE LINKE,party leader
jens spahn,613,0.219,0.688,0.093,male,0,1980,CDU,former minister
julia klöckner,280,0.304,0.632,0.064,female,1,1972,CDU,former minister
karl lauterbach,684,0.196,0.737,0.067,male,0,1963,SPD,minister
katarina barley,152,0.289,0.671,0.039,female,1,1968,SPD,former minister
klara geywitz,89,0.371,0.596,0.034,female,1,1976,SPD,minister
lars klingbeil,219,0.333,0.626,0.041,male,0,1978,SPD,party leader
malu dreyer,345,0.214,0.675,0.11,female,1,1961,SPD,minister-president
manuela schwesig,285,0.186,0.761,0.053,female,1,1974,SPD,minister-president
marco buschmann,185,0.303,0.638,0.059,male,0,1977,FDP,minister
markus söder,519,0.231,0.672,0.096,male,0,1967,CSU,minister-president
michael kretschmer,236,0.233,0.568,0.199,male,0,1975,CDU,minister-president
nancy faeser,201,0.388,0.587,0.025,female,1,1970,SPD,minister
olaf scholz,797,0.27,0.681,0.049,male,0,1958,SPD,minister
peter altmaier,190,0.279,0.668,0.053,male,0,1958,CDU,former minister
peter tschentscher,160,0.206,0.669,0.125,male,0,1966,SPD,minister-president
philipp rösler,153,0.216,0.621,0.163,male,0,1973,FDP,former party leader
reiner haseloff,149,0.268,0.705,0.027,male,0,1954,CDU,minister-president
robert habeck,413,0.293,0.654,0.053,male,0,1969,GRÜNE,minister
saskia esken,217,0.406,0.567,0.028,female,1,1961,SPD,party leader
steffi lemke,160,0.394,0.55,0.056,female,1,1968,GRÜNE,minister
stephan weil,227,0.269,0.551,0.181,male,0,1958,SPD,minister-president
susanne hennig-wellsow,81,0.333,0.63,0.037,female,1,1977,DIE LINKE,party leader
svenja schulze,207,0.391,0.531,0.077,female,1,1968,SPD,minister
tino chrupalla,160,0.275,0.656,0.069,male,0,1975,AfD,party leader
tobias hans,223,0.22,0.682,0.099,male,0,1978,CDU,minister-president
ursula von der leyen,533,0.139,0.841,0.021,female,1,1958,CDU,former minister
volker bouffier,234,0.192,0.701,0.107,male,0,1951,CDU,minister-president
volker wissing,212,0.358,0.533,0.108,male,0,1970,FDP,minister
winfried kretschmann,269,0.204,0.736,0.059,male,0,1948,GRÜNE,minister-president
wolfgang schmidt,347,0.179,0.274,0.548,male,0,1970,SPD,minister
\end{filecontents*}
\csvreader[
tabular = l|llllllll,
table head =  Name & Suggestions & Cluster 1 & Cluster 2 & Cluster 3 & Gender & Party & Birth year & Role \\\hline,
late after line = \\
]{final_pol_data.csv}
{name=\name,
google-suggestions=\googlesuggestions,
cluster-0=\clustera,
cluster-2=\clusterc,
cluster-1=\clusterb,
gender=\gender,
party=\party,
year-of-birth=\yearofbirth,
pol-role=\polrole}{%
\name & \googlesuggestions & \clustera & \clusterc & \clusterb & \gender & \party & \yearofbirth & \polrole
}%

\end{table*}

\end{document}

%% file: acronyms.tex
\acrodef{RAI}[RAI]{Recursive Algorithm Interrogation}
\acrodef{SEME}[SEME]{Search Engine Manipulation Effect}